\documentstyle[12pt]{article}
\makeatletter

\ifcase \@ptsize
\or 
\or 
\fi
\advance\textheight by \topskip

\ifcase \@ptsize
\or 
\or 
\fi

\def\section{\@startsection {section}{1}{\z@}{-3.5ex plus -1ex minus 
 -.2ex}{2.3ex plus .2ex}{\large\bf}}
\def\subsection{\@startsection{subsection}{2}{\z@}{-3.25ex plus -1ex minus 
 -.2ex}{1.5ex plus .2ex}{\sc}}
\def\@cite#1#2{\nolinebreak$^{[\scriptstyle #1\if@tempswa , #2\fi]}$}
\def\@citex[#1]#2{\if@filesw\immediate\write\@auxout{\string\citation{#2}}\fi
  \def\@citea{}\@cite{\@for\@citeb:=#2\do
    {\@citea\def\@citea{,\penalty\@m}\@ifundefined
       {b@\@citeb}{{\bf ?}\@warning
       {Citation `\@citeb' on page \thepage \space undefined}}%
{\csname b@\@citeb\endcsname}}}{#1}}
\@definecounter{footnote}

\gdef\@publabel{\hfil}
\gdef\@pubdate{\null}
\gdef\@pubnumber{\null}
\gdef\@author{\null}
\gdef\@title{\null}
\gdef\@abstract{\null}
\long\def\pubdate#1{\gdef\@pubdate{#1}}
\long\def\pubnumber#1{\gdef\@pubnumber{#1}}
\long\def\publabel#1{\gdef\@publabel{#1}}
\long\def\author#1{\gdef\@author{#1}}
\long\def\title#1{\gdef\@title{#1}}
\long\def\abstract#1{\gdef\@abstract{#1}}
\def\titlerelax{
\let\maketitle\relax
\let\settitleparameters\relax
\let\consolidatetitle\relax
\let\inittitlepage\relax
\let\finishtitlepage\relax
\let\titlepagecontents\relax
\let\multithanks\relax
\let\titlebaselines\relax
\let\@makepub\relax
\let\@maketitle\relax
\let\@makeauthor\relax
\let\@makeabstract\relax
\let\@maketitlenote\relax
\let\thanks\relax
\let\titlerelax\relax}
\def\titleclean
{\gdef\@titlenote{}
\gdef\@abstract{}
\gdef\@author{}
\gdef\@title{}
\gdef\@pubdate{}\gdef\@pubnumber{}\gdef\@publabel{}
\gdef\@dpublabel{}
}
\def\@makepub{\vbox to \z@{\hbox to \textwidth{\hfill
\@publabel \hfill
\llap{\parbox[t]{0.25\textwidth}{\raggedleft\@pubnumber}}}%
\vss}}
\def\@maketitle{\vskip 60pt \begin{center}
 {\LARGE \@title \par}
 \end{center}}
\def\@makeauthor{{\def\and{\smallskip {\normalsize \rm and\smallskip}}
\long\def\address##1{{\def\and{\\and\\}\medskip
				{\small \it \\##1\\}
}}
{\centering
 \vskip 2em
 \large \lineskip .75em
 \@author}
 \par}} 
\def\@makedate{\vskip 1.5em 
 {\raggedright \small \noindent\@pubdate \par}}
\def\@makeabstract{\vskip 1.5em
{\small 
\begin{center}
{\bf ABSTRACT\vspace{-.5em}\vspace{0pt}} 
\end{center}
\quotation \@abstract \endquotation}}
\def\maketitle{
\let\footnotesize\small \setcounter{page}{1}
\@makepub
\@maketitle
\@makeauthor
\@makeabstract
\@thanks
\@makedate
\setcounter{footnote}{0}
}

\begin{document}
\newcommand{\be}{\begin{equation}}
\newcommand{\ee}{\end{equation}}
\newcommand{\bea}{\begin{eqnarray}}
\newcommand{\eea}{\end{eqnarray}}
\newcommand{\beaa}{\begin{eqnarray*}}
\newcommand{\eeaa}{\end{eqnarray*}}

\pubnumber{hep-th/9903137}
\title{
\vspace{-0.2in}The $SO(N)$ principal chiral field on a half-line}
\author{\vspace{-0.2in}N. J. MacKay\footnote{\tt n.mackay@sheffield.ac.uk}
\address{Dept of Applied Mathematics, 
         University of Sheffield, \linebreak
         Sheffield, S3 7RH, 
         England}}
\abstract{We investigate the integrability of the $SO(N)$ principal 
chiral model on a half-line, and find that mixed Dirichlet/Neumann 
boundary conditions (as well as pure Dirichlet or Neumann) lead to 
infinitely many conserved charges classically in involution. We use
an anomaly-counting method to show that at least one non-trivial example
survives quantization, compare our results with the proposed
reflection matrices, and, based on these, make some preliminary
remarks about expected boundary bound-states.
}

\maketitle
\vspace{-0.3in}
\baselineskip 18pt
\parskip 10pt
\parindent 10pt

\section{The principal chiral field}

We first recall some preliminaries. A full treatment of the model
on the infinite line can be found elsewhere\cite{evans97}.
The principal chiral model may be defined by the lagrangian 
\be\label{pcmlagr}
{\cal L} = {\rm Tr}\left( \partial_\mu g^{-1}
 \partial^\mu g\right) \, , 
\ee
where the field $g(x^\mu)$ takes values in a compact Lie group ${\cal G}$,
here chosen to be $SO(N)$.
It has a global ${\cal G}_L\times{\cal G}_R$ symmetry 
with conserved currents 
\be\label{lrcurr} 
j(x,t)_\mu^L=\partial_{\mu} g \,g^{-1} , \qquad 
j(x,t)_\mu^R = - g^{-1}\partial_{\mu} g
\ee
which take values in the Lie algebra ${\bf g}$ of ${\cal G}$:
that is,  $j=j^a t^a$ (for $j^L$ or $j^R$: henceforth we drop this
superscript) where $t^a$ are the generators of ${\bf g}$.
The equations of motion are
\be\label{cons}
\partial^{\mu}j_{\mu}(x,t)=0 \, , \hspace{0.4in}
\partial_{\mu} j_{\nu} - \partial_{\nu} j_{\mu} - [j_{\mu},j_{\nu}] = 0 \,,
\ee
which may be combined as
\be\label{lce}
\partial_- j_+ = - \partial_+ j_- = - {1\over 2} [j_+,j_-] 
\ee
in light-cone coordinates $x^\pm = {1\over 2} (t \pm x)$.
The PCM has additional involutive discrete symmetries (`parities')
\be\label{parity}
\pi \, : \; g\mapsto g^{-1} \qquad \Rightarrow
 \qquad j^L \leftrightarrow j^R \,, 
\ee
and, for ${\cal G} = SO(N)$,
\be
\qquad\tau\,:\;g \mapsto MgM^{-1}\qquad  \Rightarrow \qquad
j^L \mapsto M j^L M^{-1} \;, \qquad
j^R \mapsto M j^R M^{-1} \,,
\ee
where $M$ is an $O(N)$ matrix with determinant $-1$.

\section{Boundary conditions for the half-line}

We place the model on the half-line $-\infty<x<0$.
Immediate suggestions for boundary conditions (BCs) might be
the (free) Neumann condition $\partial_1 g=0$ at $x=0$, implying $j_1^L(0,t)=j_1^R(0,t)=0$, or the Dirichlet condition 
$\partial_0 g=0$ at $x=0$,
implying $j_0^L(0,t)=j_0^R(0,t)=0$.
Following Moriconi and de Martino\cite{mori98},
we generalize these to mixed conditions, with (both $L$ and $R$)
$j_0(0,t)=0$ on 
$M$ components of the quantum vector multiplet, and $j_1(0,t)=0$
on the $N\!-\!M$ others. This implies
\be\label{BCs}
j_1(0,t)\in SO(M) \,,\hspace{0.5in}  j_0(0,t)\in SO(N\!-\!M) \,,
\ee
which, along with $j_\mu\rightarrow 0$ as $x\rightarrow-\infty$,
we shall take as our classical BCs. We have not attempted to 
find BCs which distinguish $L$ from $R$. Neither have we found
the boundary Lagrangian corresponding to the Dirichlet conditions.

It immediately follows from (\ref{cons}) that 
\beaa
\partial_1 j_0 (0,t) & = &  \partial_0 j_1 (0,t)  \in SO(M)\\
\partial_1 j_1 (0,t) & = &  \partial_0 j_0 (0,t)  \in SO(N\!-\!M) \,,
\eeaa
and thence that {\em all} higher derivatives of $j_0$ and $j_1$ at $x=0$
are in either the $SO(M)$ or the $SO(N\!-\!M)$ subgroup.

\section{Local conserved charges on the half-line} 

In recent work\cite{evans97} we investigated local conserved charges
on the full line, and found that the densities of the 
(odd-spin $s$) simple charges 
\be\label{lcharges}
q_s = \int_{-\infty}^\infty {\rm Tr}( j_+^{s+1}) \;,\hspace{0.5in}
q_{-s} = \int_{-\infty}^\infty {\rm Tr}( j_-^{s+1})
\ee
(which arise from the conservation laws
\be\label{laws}\left.
\partial_- {\rm Tr}(j_+^{s+1}) = 0 \;,\hspace{0.5in}
\partial_+ {\rm Tr}(j_-^{s+1}) = 0 \;\right)
\ee
must be generalized to certain polynomials 
in order to obtain a set which commute both mutually and with the 
Pfaffian charge, which exists only for $N$ even, and arises from
\be
\partial_- \left(\epsilon_{i_1 i_2 ... i_{N-1} i_N} (j_+)_{i_1i_2} ... 
(j_+)_{i_{N-1}i_N}\right)=0 \,.
\ee
We first demonstrate that our BCs lead to conservation of
\be\label{hch}
q_{|s|}\equiv q_s + q_{-s}
\ee
on the half-line.
This follows from
\bea\nonumber
{d q_{|s|}\over dt} & = & \int_{-\infty}^0 \,dx\, 
\partial_0 {\rm Tr}(j_+^{s+1}) + \partial_0 {\rm Tr}(j_-^{s+1}) \\
& = &  \int_{-\infty}^0 \,dx\, \nonumber
\partial_1 {\rm Tr}(j_+^{s+1}) - \partial_1 {\rm Tr}(j_-^{s+1}) \\
& = & {\rm Tr}\left(j(0,t)_+^{s+1}- j(0,t)_-^{s+1}\right) \,;
\label{c2}
\eea
because $s$ is odd, the trace always is always of the form
${\rm Tr}(j_0(0,t) j_1(0,t)...)$,
and so vanishes. Further, from the calculation on the
full line\cite{evans97}, we see that on the half-line
\beaa
\{q_{|r|}\,,\,q_{|s|}\} & = & {\em (const)}\; \left[{\rm Tr}(t^c j_+(0,t)^s) 
\,{\rm Tr}(t^c j_+(0,t)^r) \,-\,{\rm Tr}(t^c j_-(0,t)^s) 
\,{\rm Tr}(t^c j_-(0,t)^r) \right]\\
& = & {\em (const)} \;{\rm Tr}\left(  j_-(0,t)^{r+s} - j_+(0,t)^{r+s} \right)
\eeaa
by completeness, and so also vanishes.
The same reasoning ensures conservation and commutation of the
polynomial charges. However, it does {\em not} imply the conservation
of the Pfaffian charge, and we have found no subtler reasoning
which does so. Thus the simple charges (\ref{hch}) are the maximal
commuting set we have found.

For the group $SO(N)$ the Pfaffian charge 
is the only charge which is odd under $\tau$.
Further, if we had considered $SU(N)$, where conserved charges
on the full line exist for $s$ both odd and even, the reasoning above 
would have failed to guarantee conservation of the even-$s$ charges, 
which are precisely those odd under $\pi$. Our suspicion is therefore
that, in general, only charges which are even under all such parities
are conserved on the half-line with mixed BCs, so that boundary scattering
will mix parity-doublets. This is reminiscent of the situation in affine 
Toda field theories\cite{corr94}, where general BCs did not imply 
conservation of odd-spin charges. 

Finally we consider quantum charge conservation. Here the only method
available is the anomaly-counting of Goldschmidt and Witten\cite{gold80}, 
and the only non-trivial ({\em i.e.\ }$s>1$) charge which this method
guarantees to be conserved on the full-line, and which we have found to
be classically conserved on the half-line, is that for $s=3$.
The point is that the classical conservation law is modified
by quantum anomalies\cite{gold80,evans97} to become
\beaa
\partial_- {\rm Tr}(j_+^4) + \partial_+ {\rm Tr}(j_-^4) & = &
c_1 \left( \partial_+ {\rm Tr}(j_+\partial^2_+j_+ + {1\over 2}
j_-[j_+,\partial_+j_+]) + \partial_- {\rm Tr}(j_-\partial^2_-j_-
 + {1\over 2} j_+[j_-,\partial_-j_-]) \right) \\
&+& c_2 \left( \partial_+\left( {\rm Tr}(j_-j_+){\rm Tr}(j_+^2) \right)
+ \partial_-\left( {\rm Tr}(j_+j_-){\rm Tr}(j_-^2) \right) \right) \\
&+& c_3 \left( \partial_+ {\rm Tr}(j_-j_+^3) + \partial_-
 {\rm Tr}(j_+j_-^3) \right) \\
&+& c_4 \left( \partial_- {\rm Tr}((\partial_+j_+)^2) + 
\partial_+ {\rm Tr}((\partial_-j_-)^2) \right) \\
& + & c_5\left( \partial_- {\rm Tr}(j_+^2)^2 + \partial_+ 
{\rm Tr}(j_-^2)^2  \right)
\eeaa
(for some unknown $c_i$), the most general anomaly possible with
the correct symmetries. On the full line it is enough for conservation
that the right-hand side is a total derivative, but on the half-line 
we must check\cite{mori98} that the coefficients
of each $c_i$ individually satisfy the procedure laid out in (\ref{c2}).
They do so, and there is thus at least one non-trivial quantum
conserved charge on the half-line, which should be enough to
ensure quantum integrability. 

\section{Boundary $S$-matrices and the spectrum}

Recall that the bulk spectrum consists of particle
multiplets with masses $m_a=2m\sin\left({a\pi\over N-2}\right)$,
in representations $(V, V)$ of ${\cal G}_L\times {\cal G}_R$,
where $V$ is a reducible representation of ${\cal G}$ whose
highest component is the $a$th fundamental representation of ${\cal G}$.
These run from $a=1$ to $a=n-1$ where $N=2n+1$, 
or $a=n-2$ where $N=2n$.
There are also spinor multiplets: one in the former case, 
two in the latter, which form a $\tau$-doublet. In fact the particle
multiplets represent a larger Yangian algebra of non-local charges, 
$Y_L({\bf g}) \times Y_R({\bf g})$,
of which they are the fundamental irreducible representations.

On the half-line, if we are correct in believing that odd-parity
charges are not conserved, the parity-doublets will be broken.
Further, the charges $Q^a=\int j_0^a \,dx$ which generate 
${\cal G}_L\times {\cal G}_R$ were considered on the half-line 
by Mourad and Sasaki\cite{mour95}, who pointed out
that 
\begin{equation}\label{group}
{dQ^a\over dt}= j_1^a(0,t)\,.
\end{equation}
Those $Q^a$ corresponding to a residual 
$SO(N\!-\!M)_L\times SO(N\!-\!M)_R$ symmetry therefore remain 
conserved, as might be expected
given that the BC was free for precisely those components.
We have found no residue of the non-local charges and thus
of the Yangian symmetry, however.

Let us finish by comparing our results briefly with the reflection matrices
which, building on the work of Cherednik\cite{chere83}, 
we have constructed\cite{macka95} for the scattering of the first (vector)
and second (adjoint$\oplus$singlet) (as representations of ${\cal G}$)
multiplets of the bulk theory off the boundary. These contain a residual $SO(N\!-\!M)$ symmetry, and so might be expected to match the BCs given. 
The decomposition of the boundary $S$-matrices appears not to respect 
any residual Yangian symmetry, again as expected.

The first and second multiplet $S$-matrices take the form
\bea\nonumber
K_1(\theta) & = & \tau_1(\theta)\; (P^- - [N-2M]P^+) \\
K_2(\theta) & = & \tau_2(\theta) \;\label{K}
\left(P_A^- - [N\!-\!2M\!-\!2]P_2 +
[N\!-\!2M\!-\!2][N\!-\!2M\!+\!2] P_A^+ \right) \;,
\eea
where $\tau_1$ and $\tau_2$ are scalar prefactors\cite{macka95}
with no physical-strip poles, and 
\be 
[x]= {\theta+ix\pi/2h\over\theta-ix\pi/2h}
\ee
(with $h=2n-2$).
$P^+$ and $P^-$ are orthogonal projectors, of dimensions $M$ and $N\!-\!M$ 
respectively, taking the forms
\be
P^+=\left( \begin{array}{c|c} I_M &  L \\ \hline
 0  & 0 \end{array}\right)  \hspace{0.4in}{\rm and}\hspace{0.4in}
P^-=\left( \begin{array}{c|c} 0 &  -L \\ \hline
 0  & I_{N\!-\!M} \end{array}\right) \,,
\ee
where $L$ is an $M \!\times \!(N\!-\!M)$ matrix which cannot be determined
by the boundary YBE or the crossing/unitarity condition.
$P_A^+$ and $P_A^-$ project similarly
onto the second-rank antisymmetric tensor, whilst $P_2$ is a mixed
projector for which we have no interpretation.
To describe the principal chiral model we must use 
$$
X_1(\theta) \; K_{1\,L}(\theta)\otimes K_{1\,R}(\theta) \,,
$$
where $X_1(\theta)$ is a CDD factor with a zero at 
$\theta={N-2M\over 2h}i\pi$ (and no poles on the physical strip), 
introduced so that the overall pole
here remains simple\footnote{A suitable $X_1$ may be obtained by
setting $x=1+2M-2N$ in a minimal version (that is, with coupling-constant
dependence removed) of (3.40) of Fring and K\"oberle\cite{fring93}; 
we do not reproduce it here.}. (Notice also that if boundary scattering
really does mix parity-doublets then some more complicated construction
would be needed if we were dealing with spinor, as opposed to vector,
multiplets.)

What is $L$ for our boundary conditions? Recall our comment at the
end of section two, that at $x=0$ not only $j_0$ and $j_1$ but also all 
their (time- {\em and} space-)derivatives are in either the $SO(M)$ or the
$SO(N\!-\!M)$ subgroup. If the same is true of the boundary Lagrangian,
then there will be no operator in the model which can link
the $M$ and $N\!-\!M$ sub-multiplets, and we must have $L=0$\footnote{
If $L=0$ then we might,
following the $l\leftrightarrow n-l$ reciprocity noted\cite{doikou98}
for $SU(n)$ boundary $S$-matrices broken by diagonal boundary terms
to $SU(l)\times SU(n\!-\!l)\times U(1)$, expect an $M\leftrightarrow N\!-\!M$
symmetry. But this would place free and Dirichlet BCs on the same footing,
with residual $SO(N\!-\!M)\times SO(M)$ symmetry, contradicting (\ref{group});
further, our reflection 
matrices are not invariant under $M\leftrightarrow N\!-\!M$.}.

To understand the full pole structure, and thus the spectrum, 
is a longer-term project.
Fusing to obtain higher projectors is a difficult calculation,
and without doing so we can make no definite statements: fused
scalar prefactors can easily be vitiated by cancellations
in the matrix structure.
However, it is simple to see that $K_a$ will have a pole factor
$[N\!-\!2M\!+\!2a\!-\!2]$, and we believe that this projects onto
the $SO(M)$-restriction of the $a$th antisymmetric tensor. Following 
the ideas of Ghoshal and Zamolodchikov\cite{ghosh93}, in which a pole
at $i\theta_0$ in $K_a$ leads to a boundary bound-state (BBS)
of mass $m_a \cos\theta_0$, the BBS spectrum may therefore be 
expected, for $M<N/2$, to include states of mass
\beaa
m'_1 & = & 2 m \sin\left({\pi\over h}\right) 
\sin\left({(M\!-\!1)\pi\over h}\right), \\
&.....& \\
m_a' & = & 2m \sin\left({a\pi\over h}\right)
 \sin \left({(M\!-\!a)\pi\over h}\right), \\
&.....& \\
m'_{M\!-\!1} & = & 2 m \sin\left({(M\!-\!1)\pi\over h}\right) 
\sin\left({\pi\over h}\right) 
\eeaa
(higher than this and the poles leave the physical strip),
with the $a$th multiplet being the $a$th antisymmetric $SO(M)$ tensor.
Notice that the $a$th multiplet is degenerate in mass, and has
the same dimension as, the $(M\!-\!a)$th. As $a$ increases beyond $M>N/2$ 
the number of states falls, with a reciprocity $M\leftrightarrow
h\!-\!M$, though this does not extend to the masses, whose average increases
as $M$ increases, as seems natural for these BCs.

Finally we note that the scattering described here is off the boundary
ground state, but reflection matrices $K^{[b]}_a(\theta)$ also exist 
for the scattering of the $a$th particle off the $b$th BBS. For our
matrices (\ref{K}) we have checked explicitly, using the bootstrap
mechanism for these states, that the second BBS appears as a
pole in $K^{[1]}_1$.

{\bf Acknowledgment}\\[0.1in]
I should like to thank Patrick Dorey and Marco Moriconi for discussions.

\end{document}